\documentclass[aps,groupedaddres8s,fleqn,nofootinbib,twocolumn]{revtex4}
\usepackage{graphicx}
\usepackage{xcolor}
\usepackage{amsmath,amssymb}
\usepackage{float}
\usepackage{physics}
\usepackage{amsfonts}
\usepackage{verbatim}
\usepackage{flushend}
\usepackage{balance}
\usepackage{endnotes}
\usepackage{footnote}
\usepackage{adjustbox}
\usepackage{gensymb}
\usepackage{float}
\usepackage{epigraph}
\usepackage{xcolor}
\usepackage[utf8]{inputenc}
\usepackage{setspace}

\newcommand{\ld}{\lambda_d}

\newcommand{\kb}{k_{\mathrm{B}}}
\newcommand{\ig}{\mathrm{ig}}
\newcommand{\vib}{\mathrm{vib}}
\newcommand{\cf}{\mathrm{conf}}
\newcommand{\ex}{\mathrm{ex}}
\newcommand{\red}{\mathrm{R}}

\begin{document}
\title{Thermodynamics and Collective Modes in Hydrogen-Bonded Fluids}
\author{C. Cockrell$^{1}$ \footnote{Corresponding author. E-mail: c.j.cockrell@qmul.ac.uk}, A. Dragović$^{2}$}
\address{$^1$ Department of Materials, Imperial College London, Exhibition Road, London, SW7 2AZ, UK}
\address{$^2$  Astrophysics Group, Cavendish Laboratory, J. J. Thomson Avenue, Cambridge, CB3 0HE, UK}

%\pacs{65.20.De 65.20.JK 61.20Gy 61.20Ja}

\begin{abstract}
The thermodynamics of liquids and supercritical fluids is notorious for eluding a general theory, as can be done for crystalline solids on the basis of phonons and crystal symmetry. The extension of solid state notions such as configurational entropy and phonons to the liquid state remains an intriguing but challenging topic. This is particularly true for liquids like water whose many structural anomalies give it unique properties. Here, for simple fluids, we specify the thermodynamics across the liquid, supercritical, and gaseous states using the spectrum of propagating phonons, thereby determining the non-ideal entropy of the fluid using a single parameter arising from this phonon spectrum. This identifies a marked distinction between these ``simple" fluids, and hydrogen bonded fluids whose non-ideal entropy cannot be determined by the phonon spectrum alone. We relate this phonon theory of thermodynamics to the previously observed excess entropy scaling in liquids and how the phonon spectrum creates corresponding states across the fluid phase diagram. \textcolor{black}{Though these phenomena are closely related, there remain some differences in practice between excess entropy scaling and the similar scaling seen due to phonon thermodynamics.} These results provide important theoretical understanding to supercritical fluids, whose properties are still poorly understood despite widespread deployment in environmental and energy applications.

%Universality aids consistent understanding of physical properties and states of matter where a theory predicts how a property of a phase (solid, liquid, gas) changes with temperature or pressure. Here, we show that the matter above the critical point has a remarkable double universality. The first universality is the transition between the liquidlike and gaslike states seen in the crossover of the specific heat on the dynamical length with a fixed inversion point. The second universality is the operation of this effect in many supercritical fluids, including N$_2$, CO$_2$, Pb, H$_2$0 and Ar. Despite different structure and chemical bonding, the transition has the same fixed inversion point deep in the supercritical state. This provides new understanding of the supercritical state previously considered to be a featureless area on the phase diagram and a theoretical guide for improved deployment of supercritical fluids in green and environmental applications.
\end{abstract}

\maketitle

\section{Introduction}

%This change is limited to a finite range on the phase diagram below the critical point because the system crosses one of the three phase transition lines: melting, boiling or sublimation.
%\section{Introduction}

The properties of liquid, supercooled, and supercritical water are well known and heavily studied due to a variety of anomalies that continue to inspire enquiry\cite{Poole1992,Gallo2016}. The properties of supercritical fluids and water in particular are still poorly understood, and augmenting our understanding is critical to up-scaling the deployment of supercritical fluids in important industrial and environmental processes \cite{Kiran2000,Akiya2002,Savage1999,Huelsman2013}. The presence and nature of hydrogen bonding in water and ammonia in their supercritical states are the subject of continued debate \cite{Gorbaty1995, Schienbein2020, Kalinichev1997,Mizan1996, Cockrell2020,Eckert1996,Sarbu2000,Akiya2002,Savage1999,Huelsman2013}.

Matter beyond the liquid-gas critical point has been thought of as a homogeneous mixture of liquid and gas rather than a distinct state of matter with its own characteristics. Critical anomalies such as the Widom lines do not persist far beyond the critical point and only appear depending on the path taken through the phase diagram. An unambiguous separation of liquidlike and gaslike states which was path-independent and operated beyond the critical point was for a long time lacking.

The Frenkel line (FL) separates states which have qualitatively distinct dynamics: at temperatures below the FL particles dynamics combines diffusive jumps with oscillation and at temperatures above it is purely diffusive \cite{Brazhkin2012,Trachenko2016}. The dynamic crossover at the FL is associated with an unambiguous and path-independent transition in the supercritical state which we have termed the ``c"-transition \cite{Cockrell2022}. This transition occurs in a universal relation between the dynamics and thermodynamics across subcritical and supercritical fluid states of noble (argon), molecular (dinitrogen and carbon dioxide) and metallic (lead) compounds. Specifically, the ``c"-transition relates the isochoric heat capacity $c_V$, independently of path, to a quantity called the ``dynamial length", $\ld$ which governs the mean free path of phonons in the fluid. The turning point of the function $c_V(\ld)$ defines the transition and occurs at the universal value of $\ld = 1 $\AA \ and $c_V = 3/2 \ \kb$ (neglecting constant contributions from rotational degrees of freedom) in all the above fluids. Such a path-independent and model-free transition between liquidlike and gaslike states beyond the critical point, long thought not to exist in the supercritical state \cite{Kiran2000}, is the first of its kind.

Interestingly, this transition is not completely universal \textcolor{black}{in two ways. The first is that the transition does not operate close to the critical point where near-critical anomalies introduce path dependence and a purely phononic description becomes untenable. The second is that} the original study contained one notable exception \textcolor{black}{(throughout the phase diagram)}: water. The tetrahedral hydrogen-bonded network of water is the source of many of its anomalies, including its huge liquid heat capacity, and is the cause of water's non-conformity to the above universal transition point. The configurational contribution to the entropy defies the path-independent characterisation \textit{via} $\ld$, which arises from phononic contributions only.

In this study, we expand the portfolio of the ``c"-transition, investigating the three industrially-relevant supercritical fluids of water, ammonia and methane. We make a clear distinction between fluids which the ``c"-transition governs and those which it does not - fluids such as ammonia and water which possess strong configurational contributions to their entropy defy characterisation by $\ld$. Equivalently, governance by the ``c"-transition implies negligible configurational contributions to thermodynamic entropy even in the subcritical liquid state. We discuss the theoretical implications of the ``c"-transition as a result of this distinction, as well as its close relationship to the well-known phenomena of excess entropy scaling in liquids and supercritical fluids. \textcolor{black}{A model relating excess entropy to the vibrational properties was recently proposed \cite{Khrapak2021b}, with a focus on interactions which characterise metallic liquids, colloidal systems, and macromolecular systems. This study is therefore complementary to Ref. \cite{Khrapak2021b}, as we focus on molecular liquids and hydrogen-bonded liquids.}

\section{Methods}

\subsection{Molecular Dynamics Simulation}

We perform molecular dynamics (MD) simulations of three different molecules in this study, and compare to MD simulations performed in previous work \cite{Cockrell2021b,Cockrell2022}. For water, we use the TIP4P/2005 potential \cite{Abascal2005}, whose structural properties in the supercritical state have compared acceptably to experimental data \cite{Cockrell2020}. This is a four-site rigid model with charges plus Lennard-Jones interactions and a dipole moment. The ammonia potential we use the popular model of Impey and Klein \cite{Impey1984} which is optimised for the liquid state. Like water, this is a rigid-body potential, with five sites, charges plus Lennard-Jones interactions, and a dipole moment. For methane, we use the potential devised by Mesli, Mahboub, and Mahboub \cite{Mesli2011} which calculates the thermal, structural, and transport properties of methane well in the liquid state. This is a rigid five-site model with Buckingham interactions and no charges. The ammonia and water potentials therefore represent two classical models of hydrogen-bonded fluids, complete with electrostatic dipole moments and whose structure reasonably conforms to experimental measurements, while the methane potential lacks explicit electrostatics but has similar Van der Waals interactions and geometries to the other two. We compare these systems to nitrogen and carbon dioxide, reported previously \cite{Cockrell2022}. For nitrogen we used a two-site rigid Lennard-Jones potential \cite{Powles1976} and for carbon dioxide a rigid three-site potential with charges and Buckingham interactions and an electrostatic quadrupole moment \cite{Gao2017}. \textcolor{black}{We note that the models we consider here omit all internal vibrational degrees of freedom. Classically, these degrees of freedom each contribute a constant factor of $\kb T/2$ to the specific heat capacity. In real fluids, the contribution to the heat capacity is a function of temperature as different modes become available, making it less trivial to subtract (unlike the rotational contribution). This introduces a significant discrepancy between the heat capacity of our models and the heat capacity of the real fluids these models represent. This discrepancy is particularly pronounced for carbon dioxide and methane. However, internal bond vibrations occur at a much higher frequency and a much lower amplitude than those of the intermolecular collective modes which are the focus of this work, and therefore do not significantly couple to the phonon spectrum in small molecules. The intermolecular heat capacity (which omits rotations and internal vibrations) of a rigid model will therefore  not significantly differ from a model which incorporates molecular vibrations. In other words, the thermal discrepancy between the model systems and real systems is not expected to be significantly phononically ``active". That said, these internal degrees of freedom are not totally isolated from the rest of the system's energy and therefore will affect the phonon spectrum in principle. Internal vibrations have been to be slihtly sensitive to changes in translational molecular motion in previous experimental studies \cite{Cockrell2021b}}

Molecular dynamics simulations were performed with the DLPOLY package \textcolor{black}{version 5.0} \cite{Todorov2006}. The timestep used in all simulations was 1 fs. Electrostatics interactions were computed using the Ewald sum method with a precision of 10$^{-6}$. Equilibration of these systems\textcolor{black}{, starting from a high-temperature melt,} at each temperature and pressure state was performed in the NPT ensemble with the Berendsen thermostat and barostat for 100 picoseconds. States lay on isotherms, isobars, and isochores across the supercritical state. Each equilibrated configurations was then run in the NVE ensemble with 20 different velocity seeds for 40 picoseconds in order to generate uncorrelated initial configurations. Production runs in the NVE ensemble followed with each of these uncorrelated configurations as initial conditions. Production runs lasted for 1 ns, printing statistical data every 10 timesteps. All data which follows has been averaged over the 20 independent initial conditions at each temperature and pressure state. 

Isochoric heat capacity (per molecule), $c_V,$ was calculated in the NVE production runs using  kinetic energy fluctuations: \cite{Allen1991}:
\begin{equation}
    \label{eqn:cvnve}
    \langle K^2 \rangle - \langle K \rangle^2 = \frac{f}{2} N \kb^2 T^2 \left( 1- \frac{f \kb}{2 c_V}\right)
\end{equation}
\textcolor{black}{with $N$ the number of molecules,} $K$ the kinetic energy, and $f$ the number of translational and rotational degrees of freedom available to the molecule in question. \textcolor{black}{For monoatomic fluids (Ar) $f = 3$, for linear molecules (N$_2$, CO$_2$) $f = 5$, and for all others (CH$_4$, NH$_3$, H$_2$O) $f = 6$. None of the models we consider has internal vibrational degrees of freedom.}

The shear modulus at high frequency and shear viscosity were calculated using the molecular stress autocorrelation function, from Green-Kubo theory \cite{Zwanzig1965, Balucani1994}:
\begin{equation}
    \label{eqn:GKshearmod}
    G_\infty = \frac{V}{\kb T} \langle \sigma^{x y}(0)^2 \rangle
\end{equation}

\begin{equation}
    \label{eqn:GKviscosity}
   \eta = \frac{V}{\kb T} \int_0^{\infty} \mathrm{d} t \ \langle \sigma^{x y}(t) \sigma^{x y}(0) \rangle
\end{equation}

\noindent with $\sigma^{x y}$ an off-diagonal component of the microscopic stress tensor. The upper limit of the integral was 5 ps in practice. These functions were then averaged over each of the three independent off-diagonal components and the 20 independent initial conditions. The end result for viscosity was insensitive to adding more initial conditions or to increasing the upper limit of the integral. The (transverse) speed of sound, $c$, was then estimated as
\begin{equation}
c = \sqrt{\frac{G_{\infty}}{\rho}},
\end{equation}
with  $\rho$ the density. From here on out we use the terms $G$ and $G_{\infty}$ interchangeably.

\subsection{Specific heat and the dynamic length}

%mention something here or maybe above about ammonia, methane, water, and co2 in supercritical industries.

The relationship between thermodynamics and dynamics is manifested in the dependence of $c_V$ on the dynamic length, $\ld$. The dynamic length is calculated from the Maxwell relaxation time, $\tau$, and the speed of sound, $c$:
\begin{equation}
    \ld = c \tau.
\end{equation}
The Maxwell relaxation time measures the viscoelastic response of a medium:
\begin{equation}
    \tau = \frac{\eta}{G}.
\end{equation}
 We relate the macroscopic $\tau$ to the microscopic mean time between diffusive ``jumps" in liquid dynamics mentioned above, an assumption which is backed up by our results here and by other experimental and computational studies \cite{Jakobsen2012,Iwashita2013}. Under this assumption, the dynamic length $\ld = c \tau$ represents a lengthscale, which is the maximum wavelength of propagating transverse phonons \cite{Hansen2003}. These transverse phonons give liquids their high heat capacity and ability to support high-frequency shear stress. The decrease of $\tau$ with temperature causes $\ld$ to increase, reducing the available phase space to transverse phonons and thereby decreasing heat capacity, as the system possesses fewer degrees of freedom. In liquids this decrease of $\ld$ terminates at the boiling line, however beyond the critical point, $\ld$ continuously decreases until it becomes comparable to the mean interatomic spacing at which point no propagating transverse phonons exist at all. This also corresponds to $\tau$ becoming comparable to the mean oscillation frequency. The points on the phase diagram at which this happens constitute the dynamic definition of the FL. The loss of all transverse phonons coincides with a reduction of the heat capacity to $c_V = 2$, the thermodynamic criterion of the FL in the harmonic case. Anharmonicity causes this to change slightly such that the FL corresponds to $c_V = 2$ only approximately \cite{Trachenko2016}.

In the gaslike state above the FL, no more propagating transverse phonons exist, and $\ld$ takes on a new physical meaning. In this state motion is purely diffusive and increasing temperature increases the mean distance over which atoms travel between interactions with their neighbours. Just above the FL interaction is nearly continuous and atoms do not experience significant ballistic motion, but as temperature increases this distance coincides with the mean free path of gas theory, with interactions becoming nearly instantaneous collisions. As this mean free path increases, the system becomes self-interacting only above a certain lengthscale, which sets the minimum wavelength for the remaining longitudinal phonons. The disappearance of short wavelength collective modes as the system approaches an ideal gas is responsible for the reduction of $c_V$ from $2 \kb$ to $3 \kb/2$, the ideal gas value.

\begin{figure}
             \includegraphics[width=0.95\linewidth]{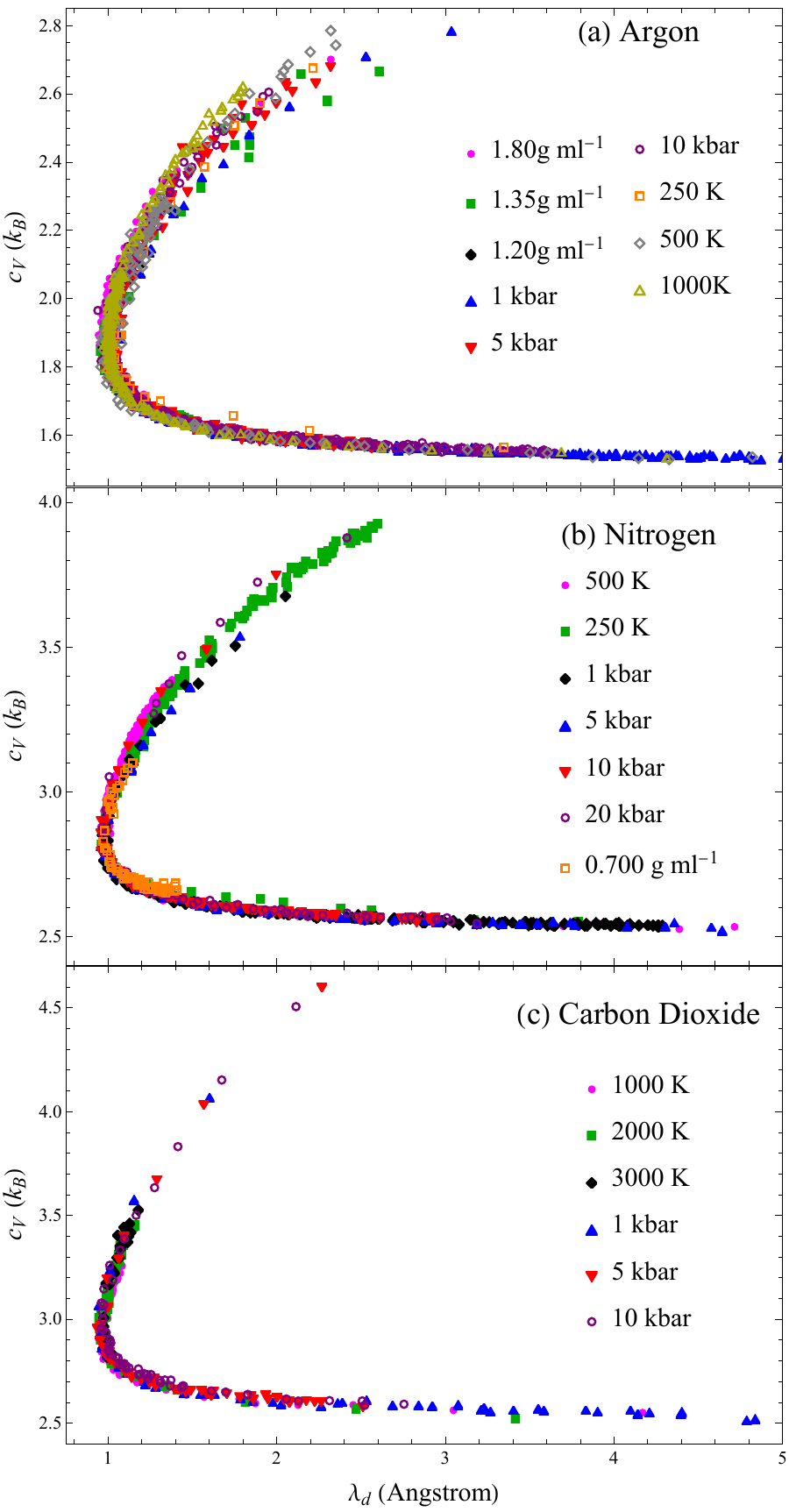}
    \caption{Heat capacity $c_V$ of (a) argon; (b) molecular nitrogen; (c) carbon dioxide as a function of the dynamic length $\ld$, showing the collapse into the path-independent ``c"-shaped curves, whose inversion point marks the transition from liquidlike to gaslike.}
    \label{fig:ljccurves}
\end{figure}

The dynamic length is therefore implicated in the evolution of $c_V$ in both the liquidlike state below the FL, where it determines the reduction of the available phase space to transverse phonons from large wavelengths downwards, and above the FL, where it determines the reduction of the available phase space to longitudinal phonons from short wavelengths upwards. In previous work \cite{Cockrell2022, Cockrell2021}, we discovered that $\ld$ entirely governs $c_V$ via a path-independent function $c_V(\ld)$ whose turning point occurs at $c_V = 1.9 \kb$, $\ld = 1$ \ \AA, corresponding to the FL. We call this curve the ``main sequence", and each fluid observing the ``c"-transition has a single main sequence curve which describes the dependence of $c_V$ on $\ld$ alone. The turning point in the main sequence is doubly universal - at every point on the supercritical phase diagram where $\ld = 1$ \ \AA in subject fluids, $c_V$ invariably equals 1.9 $\kb$ (having subtracted rotational degrees of freedom). Representative main sequence curves are displayed in Fig. \ref{fig:ljccurves} for argon and molecular nitrogen, reported previously \cite{Cockrell2022}.

\section{Results and Discussion}

\subsection{The ``c"-transition}

\begin{figure}
             \includegraphics[width=0.95\linewidth]{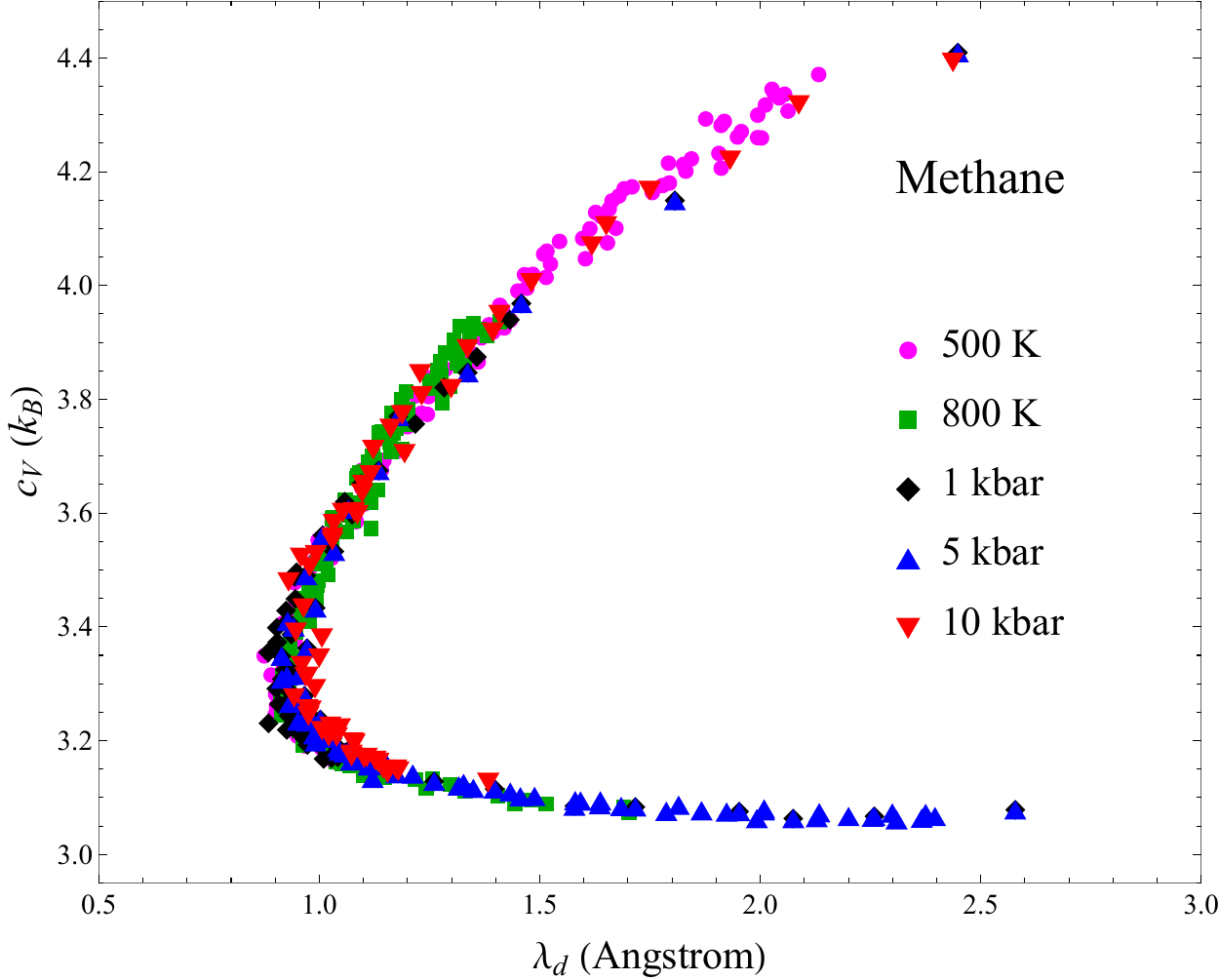}
    \caption{Heat capacity $c_V$ of methane as a function of the dynamic length $\ld$.}
    \label{fig:methaneccurves}
\end{figure}

We extend the above analysis first to methane. The model we have used for methane contains no electrostatic interactions, only Van der Waals interactions. This is appropriate as the methane molecule lacks even a quadrupole moment, becoming electrostatically interactive only at the octopole moment and higher. Methane here therefore represents a molecular with non-planar geometry, but without any significant electrostatic interactions. We plot the dependence of methane's heat capacity on the dynamic length in Fig. \ref{fig:methaneccurves}, seeing that methane is described by neat main sequence like the reference compounds in Fig. \ref{fig:ljccurves}.

\begin{figure}
             \includegraphics[width=0.95\linewidth]{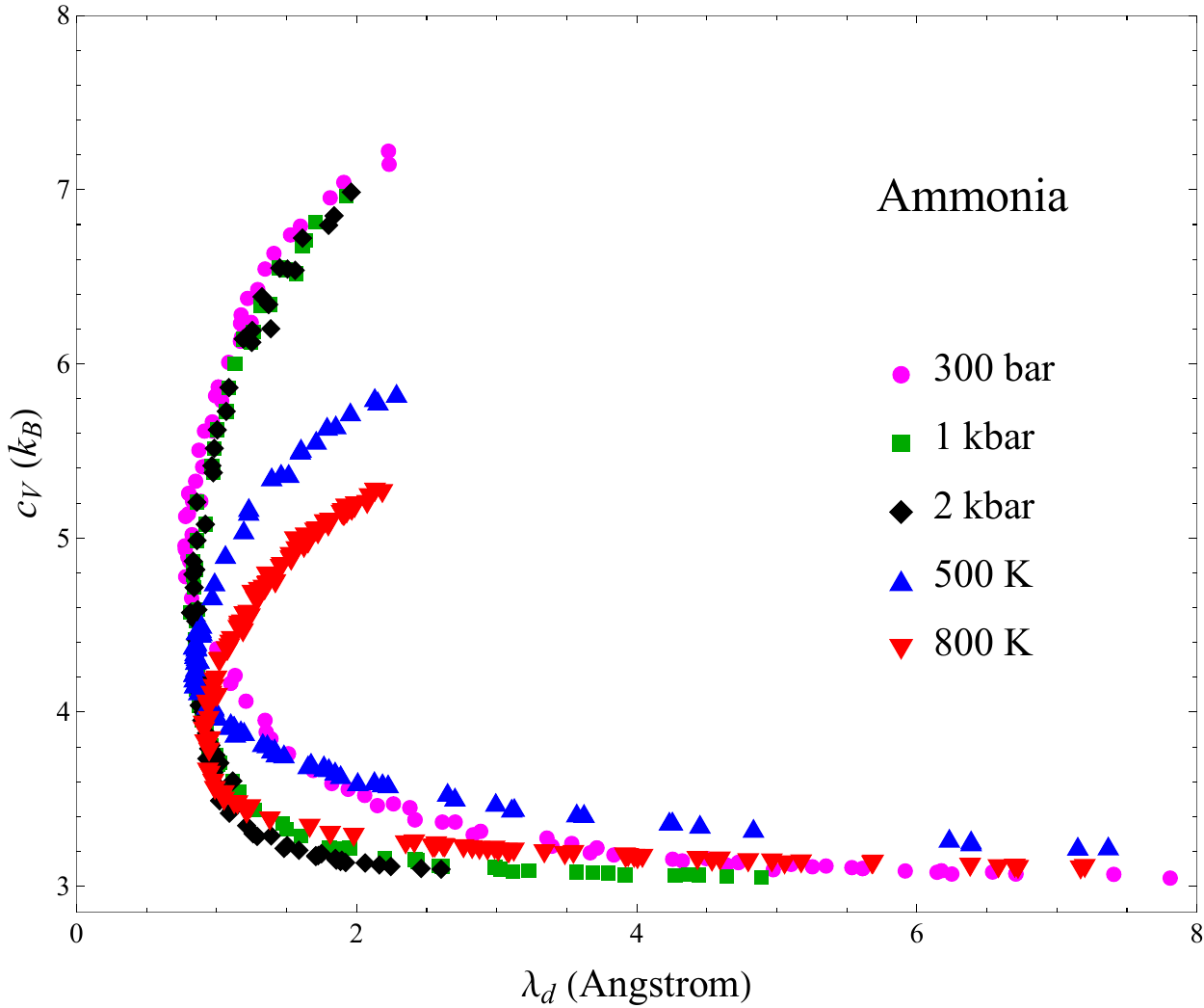}
    \caption{Heat capacity $c_V$ of ammonia as a function of the dynamic length $\ld$.}
    \label{fig:ammoniaccurves}
\end{figure}

Next we plot the dependence of $c_V$ of ammonia on $\ld$ in Fig. \ref{fig:ammoniaccurves}. Ammonia has a similar size and geometry to methane, but the ``c"-curves, $c_V(\ld)$, do not collapse onto a main sequence curve in this compound. The key difference lies in the chemistry of the nitrogen atom compared to that of the carbon atom. The electronegative nitrogen forms three bonds with hydrogen, leading to lone pairs and a dipole moment. While the turning point of each curve occurs at roughly 1 \AA, there is significant path dependence until the ``c"-curves reconcile in the ideal gas regime at $c_V = 3 \kb$.

The case is similar for water, reported previously \cite{Cockrell2022} and remade here along five isobars with better quality data. Here again there is significant path dependence, though turning points again do not stray far from 1 \AA, and the curves reconcile in the ideal gas regime.

\begin{figure}
             \includegraphics[width=0.95\linewidth]{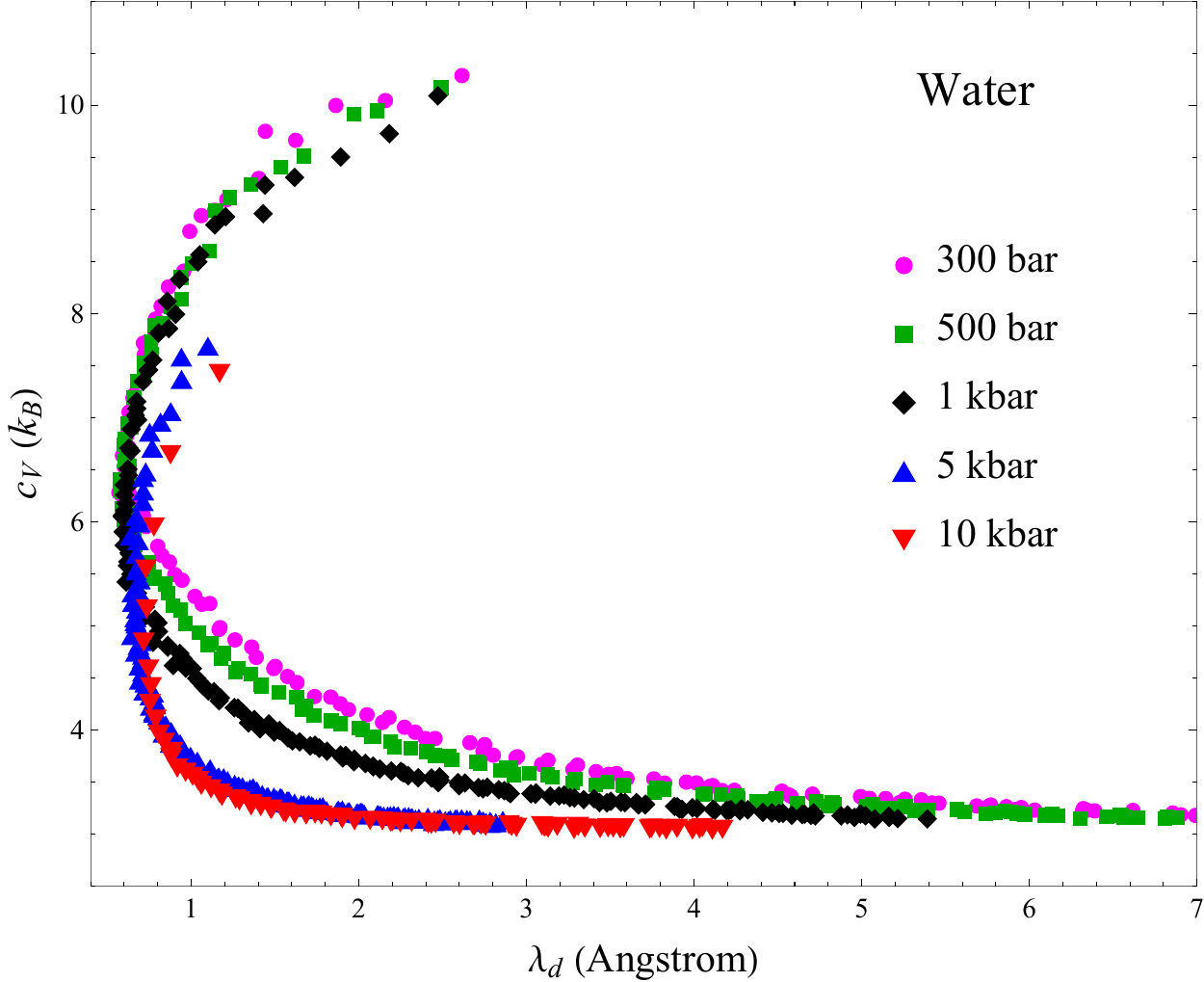}
    \caption{Heat capacity $c_V$ of water as a function of the dynamic length $\ld$.}
    \label{fig:waterccurves}
\end{figure}

We note that the presence of electrostatic interactions in the classical MD model is not the culprit behind the path-dependence. The ``c"-curves reported for carbon dioxide in Fig. \ref{fig:ljccurves} collapse onto an extremely neat main sequence, and this model possessed a quadrupole moment. What sets ammonia and water apart is the presence of hydrogen bonds in their subcritical liquid states, related to their dipole moments. The potentials used in this study recreate the structural correlations and large excess heat capacity due to hydrogen bonds The presence of hydrogen bonds in supercritical water \cite{Gorbaty1995, Schienbein2020, Kalinichev1997, Cockrell2020} and the driving mechanisms of their formation and evolution \cite{Mizan1996} in the supercritical state are contentious topics. Here we do not commit to the presence and contribution to the heat capacity of hydrogen bonds \textit{per se}, but we do note that liquids which form hydrogen bonds possess significant structural order compared to ``simple" liquids like argon, in turn giving rise to thermodynamics which may be independent of the dynamics behind $\ld$, leading to path-dependent ``c"-curves. We explore this in the following section. The reconciliation of the ``c"-curves at high temperature \textcolor{black}{(the lower branch of the ``c")} in these two anomalous fluids is indicative of thermal motion overwriting this structural order such that the heat capacity is governed almost entirely by kinetic degrees of freedom. In this case the ``c"-curves are trivially path-independent as the fluids become ideal gases and $\ld$ becomes proportional to the mean free path.

\subsection{Vibrational and Configurational Entropy}

The double universality is respected in a variety of fluids with different chemistries - argon, nitrogen, carbon dioxide, and lead. However, supercritical water, which is a polar molecule with significant configurational contributions to its entropy, does not obey this relation, demonstrating significant path dependence in the function $c_V(\ld)$. This motivates a further exploration into the thermodynamics of the ``c"-transition to ascertain what differs between fluids which observe it and those which do not.

In fluids observing the ``c"-transition, we have the universal function
\begin{equation}
    \label{eqn:universalcv}
    c_V = f(\ld),
\end{equation}
where $f$ is a function of $\ld$ only, and the dependence of $c_V$ on temperature $T$ and pressure $P$, which is generally expected for the derivative of a thermodynamic potential, is implicit with the dependence of $\ld$ on these variables. This implies that
\begin{equation}
    \label{eqn:ulambda}
    \left( \pdv{u}{T} \right)_V = f(\ld),
\end{equation}
\textcolor{black}{with $V$ the volume and $u$ the specific internal energy. With volume (and all other extensive variables) held constant in the partial derivative we can use the First Law of Thermodynamics to rewrite this in the entropy formulation of Thermodynamics \cite{Callen1985}:}
\begin{equation}
    \label{eqn:ulambda}
    \left( T\pdv{s}{T} \right)_V = f(\ld),
\end{equation}
where $s$ is the specific entropy. Rearranging, we require:
\begin{equation}
    \label{eqn:entropycondition}
    \pdv{s}{T} = \frac{f(\ld)}{T}.
\end{equation}
We postulate the decomposition of the entropy as follows:
\begin{equation}
    \label{eqn:entropydecompose}
    s = s_\ig(T,V)  + s_\vib(\ld) + s_\cf(V),
\end{equation}
where
\begin{equation}
    \label{eqn:sideal}
    s_\ig = \frac{5}{2} \kb +  \kb \log\left(\frac{V}{N} \left(\frac{m \kb T}{2 \pi \hbar}\right)^{\frac{3}{2}}\right)
\end{equation}
is the Sackur-Tetrode equation for the ideal gas entropy, $s_\vib$, the vibrational entropy, is a function of $\ld$ only. With our interpretation of $\ld$, we posit that this contribution to entropy arises solely from the fluid phonon spectrum. We nominate a residual configurational entropy, $s_\cf$, which depends on system geometry alone and therefore does not contribute to the heat capacity. The derivative of $s_\ig$ is
\begin{equation}
    \label{eqn:dsideal}
    \pdv{s_\ig}{T} = \frac{3}{2} \kb \frac{1}{T},
\end{equation}
which satisfies Eq. \ref{eqn:entropycondition}, so for the decomposition in Eq. \ref{eqn:entropydecompose} to satisfy Eq. \ref{eqn:entropycondition}, we require, via the chain rule:
\begin{equation}
    \pdv{s_\vib}{\ld} \pdv{\ld}{T} = \frac{g(\ld)}{T},
\end{equation}
where $g$ is a function of $\ld$ only. Since $s_\vib$ is also a function of $\ld$ only, we require
\begin{equation}
    \label{eqn:lambdacondition}
    \pdv{\ld}{T} = \frac{h(\ld)}{T},
\end{equation}
where, again, $h$ is a function of $\ld$ only. This is satisfied if $\ld(T, V)$ obeys a power law with $T$:
\begin{equation}
    \label{eqn:powerlaw}
    \ld(T, V) = \alpha \ |T-T_c|^{\beta},
\end{equation}
with $T_c$ the temperature of the inversion point, and the dependence on $V$ implicit with $\alpha$, $\beta$, and $T_c$. From Eq. \ref{eqn:entropydecompose}, we can write out the Helmholtz potential, $F$:
\begin{equation}
    \label{eqn:helmholtzdecompose}
    F = F_\ig(T, V) + F_\vib + F_\cf,
\end{equation}
with
\begin{equation}
    s_\vib = - \frac{1}{N} \left(\pdv{F_\vib}{T}\right)_V,
\end{equation}
where $N$ is the number of particles. To satisfy that $s_\vib$ be a function of $\ld$ only, given Eq, \ref{eqn:powerlaw}, we suggestively write (and note that a more general form is possible):
\begin{equation}
    \label{eqn:helmholtzvib}
    F_\vib(T, \ld) = - \kb T \log(Z_{\vib}(\ld)).
\end{equation}
Likewise, 
\begin{equation}
    \label{eqn:helmholtzres}
    F_\cf(T, \ld) = - \kb T \log(Z_\cf(V)).
\end{equation}
The total partition function, $Z$, is therefore
\begin{equation}
    \label{eqn:totalz}
    Z(T, V) = Q Z_\vib Z_\cf,
\end{equation}
with the kinetic part
\begin{equation}
    \label{eqn:kineticz}
    Q(T,V) = \int \prod_{i=1}^{3N} \dd p\exp \left( - \frac{p^2}{2 m \kb T} \right),
\end{equation}
where $m$ is the particle mass, and the vibrational part
\begin{equation}
    \label{eqn:configz}
    Z_\vib(\ld) Z_\cf(V) = \int \prod_{i=1}^{N} \dd^3 \mathbf{r} \exp\left(-\frac{\Phi}{ \kb T}\right),
\end{equation}
where the $\Phi$ depends on all $N$ atomic positions, and the dependence on $T$ via the integrand and $V$ via the integral limits is entirely separable into an implicit dependence with $\ld$ and a possible explicit residual dependence in $Z_\cf$. We could collapse these into one term, but it is important to remember that the heat capacity $c_V$ both temperature \textit{and volume} dependent only implicitly with $\ld$, and that any combined configurational partition function must be separable in the above manner for fluids observing the ``c"-transition.

We stop for a moment to reflect. The configurational partition function, the RHS of Eq. \ref{eqn:configz}, is generally a highly complex function of $T$, and $N$ and $V$ via integration and product limits. The kinetic partition function is easily expressed, but still depends on these variables. The dynamic length $\ld$ appears in the vibrational partition function in place of a temperature. This partition function, $Z_\vib(\ld)$, therefore describes probabilities associated with occupation of phonon modes using $\ld$ rather than $T$. This partition function counts probabilities over an abstract phase space, rather than the spatial coordinates of atoms. Considering the definition of free energy, we deduce:
\begin{equation}
    \label{eqn:freeenergy}
    F_\vib(T, \ld) = U_\vib - T S_\vib(\ld).
\end{equation}
Combining this with Eq. \ref{eqn:helmholtzvib}, we see that the internal energy corresponding to $S_\vib$ can be expressed as
\begin{equation}
    \label{eqn:energyvib}
    U_\vib(T, \ld) = \mathcal{N} \kb T,
\end{equation}
written this way to mirror the kinetic energy $\frac{3}{2} \kb N T$ with the coefficient to $\kb T$ representing the total kinetic degrees of freedom. Fluid phonons are generally anharmonic and therefore do not precisely qualify for the equipartition theorem - this $\mathcal{N}$ therefore does not necessarily represent the number of physically meaningful phonon modes in the system.

To summarise, from the path independence of $c_V$ we postulate a vibrational entropy $S_\vib$, arising from phonons, which also depends only on $\ld$. This contribution, plus the kinetic contribution $S_\ig$ and a possible residual term which depends only on the macroscopic geometry of the system, fully specifies the system entropy. Any configurational entropy, therefore, in systems observing the ``c"-transition, is entirely specified by collective modes and not by any static structural terms (as configurational entropy is commonly believed to represent). The entropic contribution of these collective modes is entirely governed by the parameter $\ld$ which represents the wavelengths available to propagating modes, such that different thermodynamic states which share a $\ld$ have identical vibrational (and therefore configurational, neglecting the residual term) entropic contributions. Corresponding to $S_\vib$ are the free energy, partition function, and internal energy, $F_\vib, Z_\vib$, and $U_\vib$. These likewise entirely express the statistical energetic contribution of the interaction potential $\Phi$ in terms of the phononic parameter $\ld$. Like how a harmonic solid can energetically and therefore entropically decomposed into collective modes \cite{Landau1969}, these fluids observing the ``c"-transition can also be fully specified by collective modes (plus kinetic contributions). In harmonic solids the phase space available to phonons does not change with thermodynamic conditions. In fluids, however, this effect is extremely important, and causes the defining feature of liquid heat capacity, namely its decrease with increasing temperature. The fluid thermodynamics therefore rests upon the parameter $\ld$, which has a straightforward qualitative interpretation and is easily and unambiguously calculable from MD simulations.

In fluids which do not observe the ``c"-transition, such as water and ammonia, the parameter $\ld$ therefore does not entirely govern the configurational entropy and free energy. In terms of the theory we laid out above, this requires the following amendment:
\begin{equation}
    \label{eqn:entropydecompose2}
    s = s_\ig(T, V) + s_\vib(\ld) + s_\cf(T, V).
\end{equation}
In other words, the residual contribution is now thermally active, as it arises not solely from the system geometry but from a microscopic structural term not implicit in the phonon spectrum. In the liquid state of water and ammonia, hydrogen bonds are unambiguously responsible for this term. In the supercritical state, the structural correlations may not qualify for the chemical definition of a hydrogen bond but the essence of the argument is unchanged. We maintain the existence of the term $s_\vib(\ld)$ despite the path dependence of $c_V$ because the function $c_V(\ld)$ is still manifested in ``c"-shaped curves which separate liquidlike and gaslike regions on a phase diagram path, meaning that the phonon contributions to heat capacity are still present, but not exhaustive. Interactions of this sort are what causes a general expression for the liquid energy / heat capacity to be impossible, as predicted by Landau \cite{Landau1980}. The path dependence present in a small degree to observant fluids like argon and methane can likewise be attributed to this non-phononic configurational term. \textcolor{black}{In real fluids, the presence of internal vibration will introduce another term. If coupling between molecular bonds and phonons is significant, the main sequence may unravel due to a dependence of the phonon spectrum on more than just $\ld$. This was seen previously \cite{Cockrell2021b} in liquid and supercritical lead, where the implicit electron-phonon coupling in the embedded atom model, rather than internal vibrations, caused slight path dependence away from the transition point at $\ld = 1.9 \kb$. For molecules where internal degrees of freedom are non-negligible, such as polymers, our model is expected to fail.}

\begin{figure}
             \includegraphics[width=0.95\linewidth]{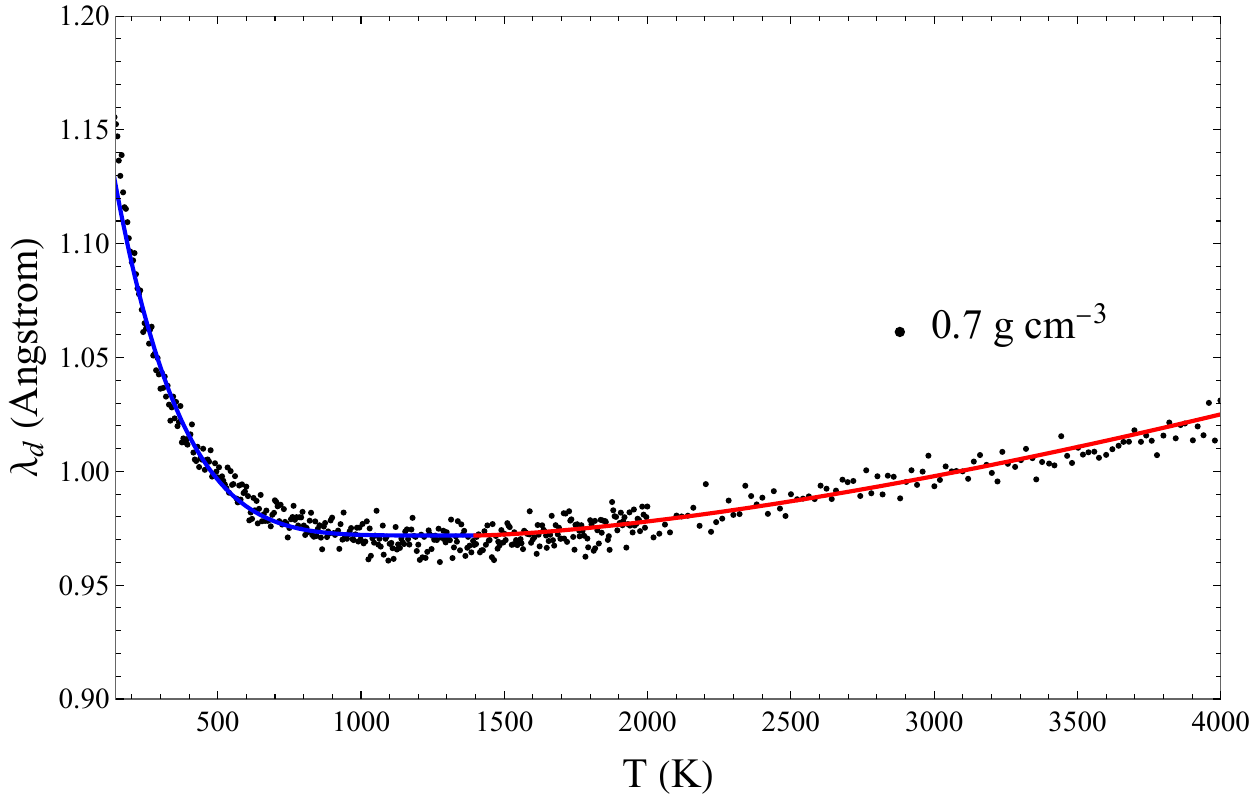}
    \caption{Dynamic length $\ld$ as a function of temperature $T$ along the 0.7 g ml$^{-1}$ isochore of $N_2$. The solid lines are power laws of the form in Eq. \ref{eqn:piecewise} fit with $T_c = 1400 $K, $d$ = 0.97 \AA, a = 1.35 $\times 10^{-18}$, $b = 5.5$, $\alpha = 5.0 \times 10^{-7}$, and $\beta = 1.5$}
    \label{fig:ldplot}
\end{figure}

We have made two interconnected assumptions in the formulation of this theory. The first is the entropy decomposition in Eq. \ref{eqn:entropydecompose}, which is not the most general way to yield Eq. \ref{eqn:universalcv}. This decomposition requires the power law dependence on temperature of the dynamic length, Eq. \ref{eqn:powerlaw}. Coming from the other direction, this power law implies the entropy decomposition. A single power law as represented in Eq. \ref{eqn:powerlaw} is not feasible, as the function $\ld(T,V)$ is quite asymmetric about its turning point $T_c$. However, because the inversion point is defined by the global minimum of $\ld$ at 1 \AA, its derivatives $\pdv{\ld}{T}$ and $\pdv{\ld}{V}$ always vanish. The function $\ld(T,V)$ can therefore acceptably be a piecewise function:
\begin{equation}
\label{eqn:piecewise}
\ld(T,V) =  \left\{
        \begin{array}{ll}
            d + a (T_c - T)^b & \quad T \leq T_c \\
            d + \alpha (T - T_c)^\beta & \quad T > T_c,
        \end{array}
    \right.
\end{equation}
with $a, b, \alpha$, and  $\beta$ encoding the volume dependence, and where $d \approx 1$ \AA. We plot $\ld$ along an isochore in supercritical N$_2$ in Fig. \ref{fig:ldplot} to assess the plausibility of this model. The fitted power laws demonstrate acceptable agreement with the curve, showing that the model is approximately correct and therefore explains the (near) path independence of $c_V(\ld)$. 

\subsection{Isomorphs and Excess Entropy Scaling}

The collapse of transport properties onto universal curves is observed along ``isomorphs" in the liquid and supercritical state \cite{Dyre2018}. Isomorphs are lines of constant excess entropy, $s_{\ex}$, along the phase diagram, where
\begin{equation}
    s_{\ex}(T, V) = s(T, V) - s_{\ig}(T, V),
\end{equation}
with $s_{\ig}$ the ideal gas entropy, as above. These lines warrant the name isomorphs insofar as the excess entropy term is accounted for by correlations the structure of the fluid. States along an isomorph differ in density and temperature, and the transport properties which undergo scaling are not ``raw", but reduced. For example, the dimensionless reduced viscosity $\eta_{\red}$ is defined
\begin{equation}
    \label{eqn:rviscosity}
    \eta_{\red} = \frac{\eta}{n^{\frac{2}{3}} (m \kb T)^{1/2}},
\end{equation}
with $n$ the molecular concentration and $m$ the molecular mass. The reduced viscosity is equal for systems \textcolor{black}{(which observe excess entropy scaling)} of the same molecular constitution and equal excess entropy. A fluid observes excess entropy scaling if its reduced transport properties are determined by the excess entropy alone. This definition mirrors that used above to define the ``c"-transition: a fluid observes the ``c"-transition if its heat capacity is governed by the dynamic length alone.

Excess entropy scaling is explained in terms of the dynamics being qualitatively scale invariant along isomorphs, a concept first introduced by Rosenfeld \cite{Rosenfeld1977}. Specifically, molecules are said to move in the same ``way" \cite{Dyre2018}, except for a uniform scaling of spatial coordinates and of time. Consideration of the dynamic length $\ld$ motivates an elaboration: equality of the motion is \textit{via} the propagating phonons present in and those disappearing from the spectrum (transverse on the upper branch, longitudinal on the lower). The dynamic length is not a dimensionless quantity, and the crossover at $c_V \approx 2$ occurs independent of density. However, the length which $\ld$ describes is a feature of collective dynamics, and its effect on the thermodynamics at best extremely weakly dependent on density, via the lengthscale at which transverse phonons cease to be.

\begin{figure}
             \includegraphics[width=0.95\linewidth]{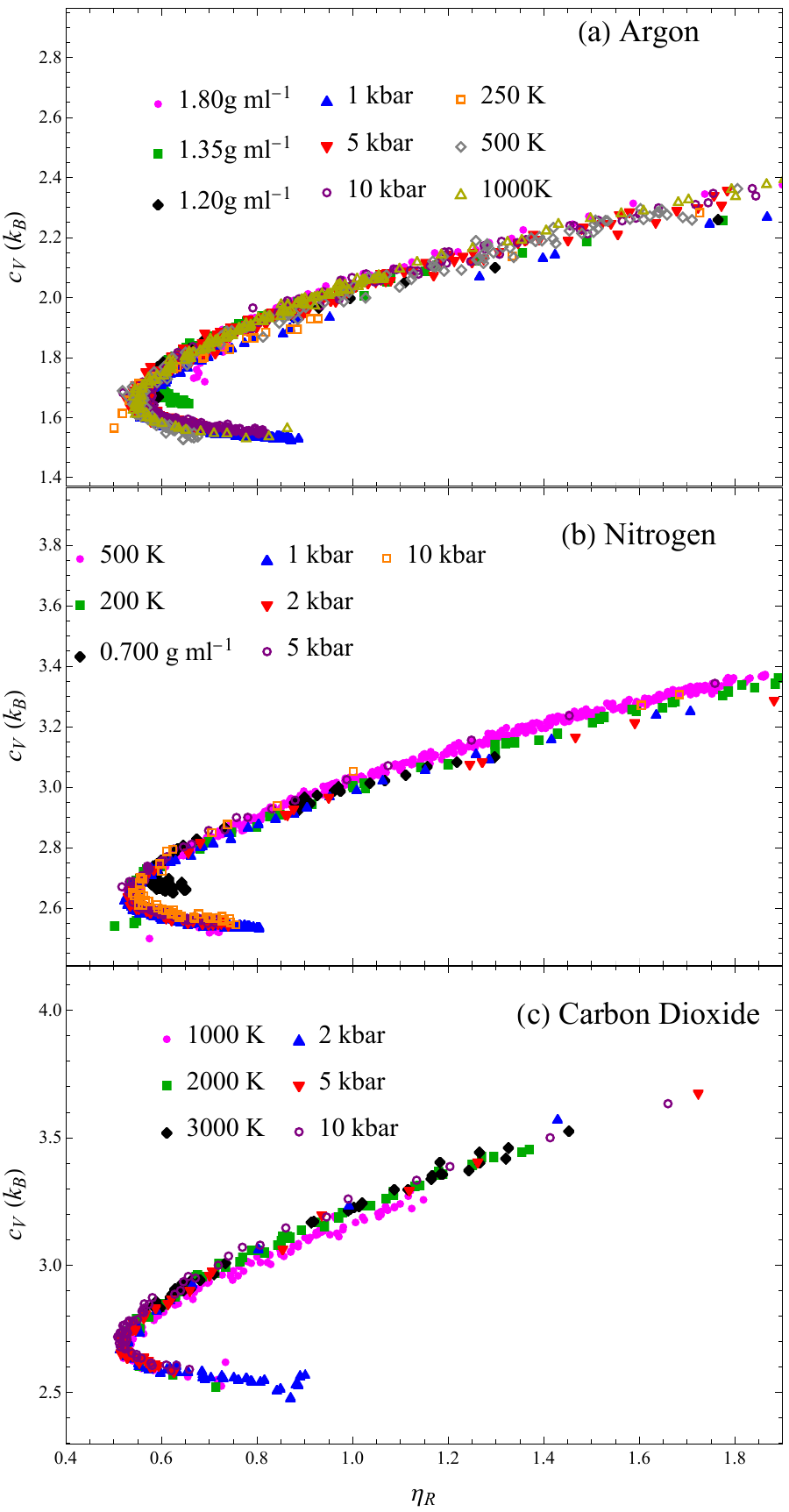}
    \caption{Heat capacity $c_V$ of (a) argon; (b) molecular nitrogen; (c) carbon dioxide as a function of the reduced viscosity $\eta_{\red}$, showing imperfect collapse beyond the liquidlike to gaslike transition.}
    \label{fig:reducedplot}
\end{figure}

In Fig. \ref{fig:reducedplot}, we plot $c_V$, having substituted the transport property $\eta_{\red}$ for the transport property $\ld$. This is a mostly acceptable substitute, demonstrating the very close relationship between the ``c"-transition and excess entropy scaling theory. Some path dependence is visible in N$_2$ and Ar at heat capacities in the gaslike state below the inversion point, an effect which was observed previously when substituting for $\ld$ in Ar \cite{Cockrell2021}. \textcolor{black}{We note that $c_V$ is not isomorph invariant, though $\ld$ is, as $\eta$, $G_{\infty}$, and $c$ each are. The quantities $\eta_\red$ and $\ld$ in particular are closely related. That said, because $c_V$ is not an isomorph invariant, isomorph theory predicts that it should not correspond directly to any invariant dynamics \cite{Bell2020}. This explains the failure of $c_V$ to collapse along $\eta_{\red}$ in Fig. \ref{fig:reducedplot}. However, the arisal of the ``c"-transition from independent considerations, initiated in Refs. \cite{Bolmatov2012}, \cite{Trachenko2016} and \cite{Cockrell2022} and elaborated upon here, suggests that $c_V$ corresponds to, at the very least, a different invariance in dynamics from that presented in isomorph theory. Isomorph theory is typically understood in fluids close to the melting point, however a recent review of dynamic models of liquid transport properties \cite{Khrapak2024} proposes a ``one-dimensional" phase diagram of fluids on the basis of excess entropy: reduced transport properties depend only on the excess entropy. In this picture, roughly speaking, dynamics are diffusion dominated when $s_{\ex} \gtrsim -1$, and vibration dominated for $s_{\ex} \lesssim -2$. Systems with the same excess entropy have the ``approximately" the same structure and dynamics in reduced units \cite{Dyre2014}. This seemingly supersedes the invariance associated with the ``c"-transition, which designates states with the same $c_V$ and $\ld$ as possessing the same phonon spectra. However there are two important caveats. The first is that, in ``c"-transition theory, phononic entropy entirely relegates configurational entropy from any independent thermodynamic meaning, whereas isomorph theory necessarily incorporates the invariant configurationa and dynamics into the excess entropy. The second is that structure and dynamics are only approximately invariant in real fluids, and the subtle differences between excess entropy scaling and the main sequence curves in $\ld$ may hint towards a subtle difference between approximate scale invariance in microscopic structure and single-particle motion and the approximate dynamic invariance in terms of collective motion (phonons). The Frenkel line, as it appears in literature \cite{Cockrell2021b}, refers to a collection of closely connected transitions which occur in the critical state. In this sense, we can comfortably state that the minimum $\ld \approx 1$ \AA \ and the point $s_{\ex} \approx -1$ correspond to the same basic physical crossover.} 

The failure to collapse for ``anomalous" systems, like water, is also well-attested in excess entropy scaling theory \cite{Fomin2010,Chopra2010,Abramson2007}, which does further suggest that similar physical mechanisms underlie these two approaches. The ``c"-transition was motivated by a theory of liquid and supercritical thermodynamics \cite{Bolmatov2012,Trachenko2016}, and its explanation of the collapse provides a comprehensive explanation of dynamics and thermodynamics in terms of collective modes, as is done in the solid state. 

Finally, we note that the heat capacity is a simpler and less ambiguous property to calculate in MD simulations than entropy, making it an attractive alternative when considering the collapse of transport properties. \textcolor{black}{The recently developed Freezing Density Scaling theory \cite{Khrapak2022c} also addresses this computational complication by substituting $n/n_{\mathrm{fr}}$ for $s_{\ex}$ along an isotherm, where $n_{\mathrm{fr}}$ is the concentration at the melting point. The fluids in which this substitution is acceptable should roughly correspond to those who observe the ``c"-transition.}

\section{Conclusions}

Configurational and vibrational entropy are extremely powerful concepts when applied to crystalline solids where phononic and elastic properties are deducible from the lattice structure itself. In the fluid state discerning and discriminating these contributions is less straightforward. We have demonstrated that the path-independent main sequence of the ``c"-curves implies that a configurational entropy term, distinct from a vibrational entropy which arises from collective excitations, is unnecessary to describe the system thermodynamics. This is not to say that local intermolecular structure does not beget energetic degrees of freedom, but instead that this structural contribution to entropy is entirely encoded within the spectrum of collective excitations. In other words, the specifics of local intermolecular structure contribute to thermodynamics only \textit{via} their participation in attenuating phonons, characterised by $\ld$ alone. The shape of the main sequence, we note, is still system-dependent, even though the inversion point is universal.

The idea of a configurational entropy, in the sense that it is used in the solid state, is only non-redundant in systems which do not observe the ``c"-transition. Such systems are predictably those with strong and orientation-dependent interactions which produce additional degrees of freedom, the evolution of which with temperature is not fully captured by the isotropic parameter $\ld$. 

The nature of the transition at the inversion point is mysterious. We have posited that $c_V$ is a single analytic function of $\ld$ across its entire domain, however the theory undergoes no essential changes if we allow piecewise function, as in Eq. \ref{eqn:piecewise} for $\ld(T,V)$. The function and its derivatives, however, appear completely continuous at the inversion point. The two branches of the main sequence, centred on the universal minimum of $\ld = $1 \AA, represent the two different dynamic-thermodynamic regimes below and above the FL. The interpretation of $\ld$ becomes murky around this transition, as the propagation length of transverse phonons reduces to the interatomic separation and the mean free path of atom starts to exceed the interatomic separation. Nonetheless, the universal dependence of $c_V$ on $\ld$ in this transition region strongly suggests that $\ld$ remains physically meaningful. The universal main sequence arises from a thermodynamic entropy which depends on a dynamic parameter $\ld$. The branches of the main sequence meet at the universal minimum in $\ld$, separating two dynamically distinct regimes. On the basis of these observations we denote the ``c"-transition as a continuous dynamic-thermodynamic transition which operates across the supercritical state of all ``simple fluids". \textcolor{black}{This picture is tantalisingly close to the universal description of simple fluids which arises from isomorph theory \cite{Khrapak2024}, though the relationship between excess entropy and reduced transport properties and between heat capacity and the dynamic length do not quite line up. This may be due to the difference between single-particle motion and collective motion implicated in isomorph theory and ``c"-transition theory respectively.}

The excellent solvation properties of supercritical fluids depends on the combination of their strong intermolecular interactions and high density with high diffusivity. In this sense they are the ``best of both worlds", referring to liquids and gases. The universal inversion point represents the total mixing of liquidlike and gaslike properties, where short-range elasticity which inhibits molecular mobility is minimised but where high density and local order maintain strong intermolecular interactions. Solubility maxima (known as ``ridges'') and optimal extracting and dissolving abilities of supercritical fluids have been observed to coincide with the FL \cite{Wang2017,Cockrell2021b}. Better theoretical understanding of the supercritical state is a crucial component to scaling up existing applications \cite{Eckert1996,Sarbu2000,Kiran2000}. The ``c"-transition is therefore a valuable guide which identifies a line on the phase diagram where this liquid-gas mixing is maximised. 

%In summary, we have shown that the supercritical state has a remarkable double universality. First, the transition between the liquidlike and gaslike states is characterised by fixed inversion point and near path-independence. Second, this effect universally applies to many supercritical fluids. This provides new understanding of the supercritical state of matter and a theoretical guide for improved deployment of supercritical fluids in green and environmental applications.

\section{Acknowledgements}

%We are grateful to V. V. Brazhkin and J. Proctor for discussions.

%Author contributions: the authors have contributed equally to this paper.

%Competing interests: The authors declare that they have no competing interests.

%Data and materials availability: All data needed to evaluate the conclusions in the paper are present in the paper. 

DL$\_$POLY is open source available under LGPL 3.0 developed mainly at Science and Technology Facilities Council Daresbury Laboratory and is freely downloadable at gitlab.com/ccp5/dl-poly. This research utilized Queen Mary’s MidPlus computational facilities, supported by QMUL Research-IT.

\bibliography{collection_2024}
\bibliographystyle{unsrt}

\end{document}